\documentclass[english,pra,showpacs,showkeys,tightenlines,secnumarabic,11pt]{revtex4}
\usepackage[T1]{fontenc}
\usepackage[latin1]{inputenc}
\usepackage{amsmath}
\usepackage{graphicx}
\usepackage{amssymb}
\usepackage{epsfig}
\usepackage{graphics}
\usepackage[mathscr]{euscript}
\usepackage{psfrag}
\usepackage{pstricks}
\usepackage{pst-node}
\usepackage{babel}
\newcommand{\bbeta}{\mbox{\boldmath$\beta$}}
\newcommand{\esp}[1]{\, e^{\,\,\textstyle {#1}}}

\begin{document}
\title{Inclusive and "exclusive" cross sections in the regime of multiple parton collisions}
\author{G. Calucci}
\email{giorgio.calucci@ts.infn.it}
\author{D. Treleani}
\email{daniele.treleani@ts.infn.it} \affiliation{ Dipartimento di
Fisica Teorica dell'Universit\`a di Trieste and INFN, Sezione di
Trieste,\\ Strada Costiera 11, Miramare-Grignano, I-34014 Trieste,
Italy.}
\begin{abstract}
The large luminosity and the increased kinematical domain
accessible at the LHC will allow to isolate large numbers of
events with multiple parton collisions. Interestingly, the hadron
is probed in different points simultaneously in the process, which
allows to obtain information on the correlations between partons
in the hadron structure. The whole experimentally accessible
information on multiple parton collisions consists in the
probability distribution of the elementary interactions, while the
inclusive cross sections, usually considered in large $p_t$
processes, acquire a precise statistical meaning as moments of the
multiplicity distribution. Conversely different and more
"exclusive" cross sections become experimentally viable, providing
complementary information on the interaction dynamics. The matter
is discussed in details in the case of hadron-hadron collisions
and the role of parton correlations is outlined both in the
inclusive and in the "exclusive" cross sections.
\end{abstract}



 \maketitle

\section{Introduction}

The overall features of the minimum bias and of the underlying
event in hard hadronic interactions cannot be reproduced without
implementing multiple parton interactions (MPI) in the event
generator\cite{Sjostrand:1987su}\cite{Acosta:2004wqa}\cite{Sjostrand:2006za}\cite{Bahr:2008dy}.
On the other hand, the direct measurement of MPI requires
identifying the final fragments of the multiple process, to
reconstruct the kinematics at the parton level. The sizable
reduction of statistics, caused by the request of large momentum
exchange in each hard interaction, has restricted considerably up
to now the possibilities of a direct study of the phenomenon and
measurements have been possible only in very limited phase space
intervals\cite{Akesson:1986iv}\cite{Alitti:1991rd}\cite{Abe:1997bp}\cite{Abe:1997xk}.
The large energy in the $pp$ c.m. system may however justify the
expectation that, at the Large Hadron Collider (LHC), MPI will not
be important only to describe the properties of the minimum bias
and of the underlying event, giving a significant contribution
also, as an example, to the cross section to produce jets with
large $p_t$.

A preliminary requirement for the study of MPI in large $p_t$ jet
production is the separation of the contribution of multiple
radiation, as the two mechanisms can produce the same final state.
When looking at the properties of the minimum bias and of the
underlying event the separation between the two sources is rather
ambiguous. In the case of direct observation of the large $p_t$
scattered partons, one may however expect that the separation of
the contribution of MPI form the leading single scattering
perturbative Quantum Cromo Dynamics (QCD) mechanism might be done
more confidently. On one side hard radiation induces an azimuthal
correlation between the produced jets, which is absent in the case
of jets produced through MPI, on the other the two mechanisms have
a very different dependence on the lower value of the transverse
momentum introduced to observe the jets and on the c.m. energy,
being the MPI a power correction to the leading QCD single
scattering mechanism. Azimuthal correlations were in fact
successfully used by CDF to isolate and measure double parton
collisions at Fermilab and the statement made by the experiment,
that some 17\% of their sample of double scattering events was
contaminated by triple scatterings\cite{Abe:1997xk}, justifies the
hope that the larger energy available at the LHC will allow to
isolate events with triple and perhaps also with quadruple parton
interactions without much ambiguity.

One should nevertheless point out that, while in the case of
double collisions the separation of hard radiation and of the MPI
contribution may be relatively simple, the best strategy to
isolate the contribution of multiple collisions, when the number
of scatterings is large, may not be readily identified. Although
it might still be relatively simple, at least for triple
scatterings, in the case of particular channels, like W production
plus jets or Drell-Yan plus jets, with the jets in particular
kinematical configurations. In more general cases, the strategy to
separate the contribution of MPI from the contribution of the
leading single QCD production mechanism deserves a dedicated study
and will not be addressed in the present paper. Rather we will
assume that the increased parton luminosity and the sizably higher
value of the momentum exchanged in a typical partonic interaction
at the LHC will allow the direct identification of a sizable
number of events, not only with double but also with triple and
perhaps quadruple parton interactions.

Interestingly, one may expect that a detailed study of the phenomenon will allow
to access several features of the hadron structure still
unexplored and most likely unexpected. The anomalously small value
of the scale factor, characterizing the measurement of double
parton interactions by CDF at
Tevatron\cite{Abe:1997bp}\cite{Abe:1997xk}\cite{Calucci:1997ii},
and the need to use a parton density much different as compared
with the hadron form factor, in the actual Montecarlo
generators\cite{Sjostrand:1987su}\cite{Sjostrand:2006za}, are
indications that unpredicted features of the hadronic structure
may play a non secondary role in the phenomenon.

A general frame to identify and organize the novel information in
a systematic way is hence highly desirable. The present paper
represents an attempt to provide such a frame. One should stress
that our goal is by no means to provide a scheme for a general
description of the overall inelastic event, so our aim is very
different in comparison with Montecarlo models. Our final states
consists only of the large $p_t$ partons generated by MPI, where
we assume that one will be able to recognize and count the number
of interactions in some phase space windows. In addition, we are
interested only in a small fraction of all inelastic events and
only in few particular final state observables (like for example
the jet's momenta). Our aim is to show how such quantities are
linked to the correlations of the partons in the hadron structure,
while remaining as general as possible. So we will not attempt to
compare various ideas about correlations, which have been
explicitly implemented, in particular in Montecarlo models. Rather
we will try to identify observables which have a direct link with
correlations and to establish sum rules which, allowing to control
the consistency of the scheme at a given order in the number of
MPI, will help to disentangle the effects of correlations from the
effects of unitarization.

A preliminary comment concerns the hard cross sections accessible in
the regime of MPI. As it will be discussed
in the next sections, in the case of hadron-hadron interactions
the integrated inclusive cross sections of the QCD-parton model
are essentially the moments of the distribution in the number of
collisions. The whole information of a probabilistic distribution
is either given by the set of the different terms of the
distribution or, alternatively, by the set of its moments. One may
hence consider two different sets of cross sections, the inclusive
cross sections, given by the moments of the distribution, and the
"exclusive" cross sections, given by the probabilities of the
different terms of the distribution in the number of collisions.

The two different sets of cross sections correspond to two
different criteria of selection of events. In the case of the
inclusive cross section of a $K$-parton scattering, one needs to
collect all events where there are {\it at least} $K$-parton
collisions in a given final state phase space interval. All events
with $N$-parton collisions, where $N\ge K$, hence contribute with
a multiplicity factor $N\choose{K}$. On the other hand the
"exclusive" cross section of a $K$-parton scattering process is
measured by selecting all events with {\it only} $K$-parton
collisions in the phase space interval of interest. The cross
sections called here "exclusive" are in fact partially inclusive
cross sections, since one sums over all partons outside the given
phase space interval and on the soft fragments.

By measuring the inclusive cross sections, which are
related to the moments of the distribution, one has the advantage
of approaching the problem in a way ordered in complexity. In the
case of the "exclusive" cross sections, each term is on the
contrary related to the whole set of the multiple scattering
series, since each term contains the probability of no
interaction, which carries information on the whole series of
multiple collisions. In other words each "exclusive" cross section
requires, at least in principle, the knowledge of an infinite set
of non perturbative inputs to be evaluated. On the other hand the
"exclusive" cross sections are well defined experimental
quantities, characterizing the regime of hard interactions. It is
hence worth trying to discuss also the "exclusive" cross sections
in the framework of perturbation theory.

The simplest case of "exclusive" cross section is the hard cross
section itself, defined as the contributions to the inelastic
cross section of all events with at least one hard partonic
interaction. In the case of hadron-hadron collisions all terms
with correlations can be re-summed, when only
two-body parton correlations are taken into account\cite{Calucci:1997ii}. As a
result of resummation the effects of the correlation terms is
rather indirect in the hard cross section, in such a way that the
information on parton correlations cannot be disentangled in a
simple way. Rather obviously a similar
property holds for all "exclusive" cross sections.

Isolating and organizing the correlation terms in order of growing
importance is nevertheless possible also in the case of the
"exclusive" cross sections. As a consequence of incoherence
between the different partonic interactions, MPI
are in fact described by a (different) probabilistic distribution
in each phase space interval and one may reduce or increase the
average number of partonic collisions by restricting or enlarging
the interval in transverse momenta and in rapidity where final
states are observed. It is hence always possible to give an
operational meaning to an expansion in the number of collisions.
The expansion of the "exclusive" cross sections in number of
elementary collisions allows to isolate the different correlation
terms and to obtain interesting consistency relations between
inclusive and "exclusive" cross sections.

The paper is organized as follows. After recalling the main
features of the simplest Poissonian model, which
represents the basis of most of the practical implementations of multiple
parton collisions, we will generalize the approach by including parton correlations in the picture
of the interaction through a functional approach. Keeping only two-body parton correlations into
account, explicit expressions will hence be obtained both for inclusive
and "exclusive" cross sections, in the latter case by expanding in
the number of parton collisions. A simple and experimentally
testable connection between the two quantities, at the lowest non
trivial order in the number of collisions, will be finally
derived.

\section{The simplest Poissonian model}

A standard way to introduce MPI in
Montecarlo codes is to assume a Poissonian distribution of
multiple parton collisions, with  average number depending on the
value of the impact parameter. The approach was introduced long
ago\cite{Capella:1986cm}\cite{Sjostrand:1987su}\cite{Ametller:1987ru}
and critically re-discussed recently\cite{Rogers:2008ua}. Of course the Poissonian is only a starting point for the Montecarlo generators, which are in fact much more elaborated. Both the JIMMY\cite{Butterworth:1996zw} and PYTHIA\cite{Sjostrand:1987su} include
momentum correlations and the newest variant of PYTHIA\cite{Sjostrand:2008vc}
imposes also flavor conservation. In the present section we discuss the simplest option, namely the case where no correlation is taken into account in a hadronic interaction at a given value of the impact parameter. Correlations will be introduced in a rather general way in the following section.

One may start by
introducing the three dimensional parton density $D(x,b)$, namely
the average number of partons with a given momentum fraction $x$
and with transverse coordinate $b$ (the dependence on flavor and
on the resolution of the process is understood) and one makes the
simplifying assumption $D(x,b)=G(x)f(b)$, with $G(x)$ the usual
parton distribution function and $f(b)$ normalized to one. The
inclusive cross section for large $p_t$ parton production
$\sigma_S$ may hence be expressed as

\begin{eqnarray}
\sigma_S=\int_{p_t^c}G(x)\hat{\sigma}(x,x')G(x') dxdx'=\int_{p_t^c}G(x)f(b) \hat{\sigma}(x,x')G(x')f(b-\beta)
         d^2 bd^2\beta dxdx'
\end{eqnarray}

\noindent where $p_t^c$ is a cutoff introduced to distinguish hard
and soft parton collisions and $\beta$ the hadronic impact
parameter. The expression allows a simple geometrical
interpretation. Given the large momentum exchange, which localizes
the partonic interaction inside the overlap volume of the two
hadrons, one may identify with $b$ and $b-\beta$ the transverse
coordinates of the two colliding partons while $\beta$ is the
impact parameter of the hadronic collision.

\noindent Neglecting all correlations in the multi-parton
distributions, the inclusive cross section for a double parton
scattering $\sigma_D$ is analogously expressed by

\begin{eqnarray}
\sigma_D&=&{1\over 2!}\int_{p_t^c}G(x_1)f(b_1)\hat{\sigma}(x_1,x_1')G(x_1')f(b_1-\beta)d^2b_1dx_1dx_1'\times\nonumber\\
&&\qquad\qquad\times
         G(x_2)f(b_2)\hat{\sigma}(x_2,x_2') G(x_2')f(b_2-\beta)
         d^2b_2dx_2dx_2'd^2\beta\nonumber\\
         &=&\int{1\over 2!}\Big(\int_{p_t^c} G(x)f(b)\hat{\sigma}(x,x')G(x')f(b-\beta)d^2bdxdx'\Big)^2d^2\beta
\end{eqnarray}

\noindent which allows an analogous geometrical interpretation.
The expression may be readily generalized to the case of the
inclusive cross section for $N$-parton scatterings $\sigma_N$:

\begin{eqnarray}
\sigma_N=\int{1\over N!}\Big(\int_{p_t^c}G(x)f(b)\hat{\sigma}(x,x')G(x')f(b-\beta)d^2bdxdx'\Big)^Nd^2\beta
\end{eqnarray}

The cross sections are divergent for $p_t^c\to0$. The unitarity
problem is solved by normalizing the integrand which, being
dimensionless, may be understood as the probability to have a
$N$th parton collision process in a inelastic event. One may hence
introduce $P_N(\beta)$, the probability of having $N$ parton
collisions in a hadronic interaction at impact parameter $\beta$:

\begin{equation}
P_N(\beta)\equiv{\bigl(\sigma_SF(\beta)\bigr)^N\over N!}
 e^{-\sigma_SF(\beta)},\quad\int_{p_t^c}G(x)f(b)\hat{\sigma}(x,x')G(x')f(b-\beta)d^2bdxdx'\equiv \sigma_SF(\beta)\end{equation}

By summing all probabilities one obtains the hard cross section
$\sigma_{hard}$, namely the contribution to the inelastic cross
section due to all events with {\it at least} one parton collision
with momentum transfer greater than the cutoff $p_t^c$:

\begin{eqnarray}
\sigma_{hard}=\sum_{N=1}^{\infty}\int P_N(\beta)d^2\beta =\sum_{N=1}^{\infty}\int d^2\beta{\bigl(\sigma_SF(\beta)\bigr)^N\over N!}
 e^{-\sigma_SF(\beta)}=\int d^2\beta\Bigl[1-e^{-\sigma_SF(\beta)}\Bigr]
\end{eqnarray}

\noindent Notice that
$\sigma_{hard}$ is finite in the infrared limit, which allows to
express the inelastic cross section as $\sigma_{inel}=\sigma_{soft}+\sigma_{hard}$ with $\sigma_{soft}$ the soft contribution, the two
terms $\sigma_{soft}$ and $\sigma_{hard}$ being defined
through the cutoff in the momentum exchanged at parton level,
$p_t^c$.

As mentioned the inclusive cross sections are related to
the moments of the distribution. The property is immediate in the
case of the Poissonian model. Working out the average number of
collisions one in fact obtains:

\begin{eqnarray}
\langle N\rangle\sigma_{hard}=\int d^2\beta\sum_{N=1}^{\infty}NP_N(\beta)=\int d^2\beta\sum_{N=1}^{\infty}
{N\bigl[\sigma_SF(\beta)\bigr]^N\over N!}
 e^{-\sigma_SF(\beta)}
=\int d^2\beta \sigma_SF(\beta)=\sigma_S
\end{eqnarray}

\noindent which is the single scattering inclusive cross section,
while, more in general, one may write:

\begin{eqnarray}
{\langle N(N-1)\dots(N-K+1)\rangle\over K!}\sigma_{hard}
&=&\int d^2\beta \sum_{N=1}^{\infty}
{N(N-1)\dots(N-K+1)\over K!}P_N(\beta)\nonumber\\
&=&\int d^2\beta {1\over K!}\bigl[\sigma_SF(\beta)\bigr]^K=\sigma_K
\end{eqnarray}

A few features of the model are worth to be pointed out.

\noindent - The cross sections of the two sets, inclusive and
"exclusive" (the latter being the contributions to
$\sigma_{hard}$ due to the different numbers of collisions) are expressed fully explicitly in terms of the same quantity, the average number of collisions at a fixed impact parameter $\sigma_SF(\beta)$. In a more general case, when correlations are taken into account, the two sets of cross sections can provide a complementary information on the multi parton structure of the hadron.

\noindent - All inclusive cross sections are divergent in the
infrared limit. The relation with the multiplicity of collisions
shows that the divergence is due to the moments. In other words the cause of the divergence is the number of collisions, which become very
large at low $p_t$.

\noindent - While all inclusive cross sections become increasingly
large at low $p_t$, due to the increasingly large number of parton
interactions, all "exclusive" cross section where the number of
participating hard partons is kept fixed become, on the contrary,
smaller and smaller for the same reason.

\noindent - In the model one assumes that the transverse hadron
size is the same in each interaction, so one is identifying the
hadron with its average configuration. Hadronic diffraction shows
however that hadron fluctuations are not a negligible effect also
at LHC energies\cite{Blaettel:1993ah}\cite{Gotsman:2007ac}. It was
moreover pointed out that, most likely, the importance of hadronic
fluctuations grows when the number of partonic collisions is
large\cite{Frankfurt:2008vi}\cite{Treleani:2008hq}.

\section{A functional approach}

\par MPI can be discussed in very general terms by a functional
formalism\cite{Calucci:1991qq}\cite{Calucci:1997ii}.

Let  $W_n(u_1\dots u_n)$ be the exclusive $n$-parton
distributions, namely the probabilities to have the hadron in a
configuration with $n$ partons with coordinates $u_i$, which
represent the variables $(b_i,x_i)$, being $b$ the transverse
partonic coordinate and $x$ the corresponding fractional momentum.
The scale for the distributions is given by the cut off
$p_t^{min}$ that defines the separation between soft and hard
collisions. The distributions are symmetric in the variables
$u_i$.

\noindent One may introduce the generating
functional:

\begin{equation}{\cal Z}[J]=\sum_n{1\over n!}\int J(u_1)\dots J(u_n)W_n(u_1\dots u_n)
du_1\dots du_n.\end{equation}

\noindent The conservation of the probability implies the
normalization condition ${\cal Z}[1]=1$.

\noindent The probabilities of the various configurations, namely the  exclusive
distributions, are the coefficients of the expansion of ${\cal
Z}[J]$ for $J=0$. The coefficients of the expansion of ${\cal
Z}[J]$ for $J=1$ give the many body densities, i.e. the inclusive
distributions:

\begin{eqnarray}D_1(u)={\delta{\cal Z}\over \delta J(u)}\biggm|_{J=1},\quad
                 \quad\
     D_2(u_1,u_2)={\delta^2{\cal Z}\over \delta J(u_1)\delta J(u_2)}
                  \biggm|_{J=1}\quad\dots\end{eqnarray}

\noindent Correlations, which describe how much the distribution deviates from a
Poissonian, are obtained by the expansion of the logarithm of the generating
functional, ${\cal F}[J]\equiv{\rm ln}{\cal Z}[J]$, for $J=1$:

\begin{eqnarray}{\cal F}[J]=\int D_1(u)[J(u)-1]du+\sum_{n=2}^{\infty}{1\over n!}
\int C_n(u_1&\dots& u_n)\bigl[J(u_1)-1\bigr]\dots\cr
  &\dots&\bigl[J(u_n)-1\bigr]
du_1\dots du_n\end{eqnarray}

\noindent  Obviously one has ${\cal F}[1]=0$ and, in the
Poissonian case, $C_n\equiv 0, n\ge 2$.

\noindent  Given the general
expressions for the multiparton distributions one can write down
an expression for the semi-hard cross section which takes into
account of all possible MPI:

\begin{eqnarray}\sigma_{hard}=\int d\beta\int&&\sum_n{1\over n!}
  {\delta\over \delta J(u_1)}\dots
  {\delta\over \delta J(u_n)}{\cal Z}_A[J]\nonumber\\
  &\times&\sum_m{1\over m!}
  {\delta\over \delta J'(u_1'-\beta)}\dots
  {\delta\over \delta J'(u_m'-\beta)}{\cal Z}_B[J']\cr
&\times&\Bigl\{1-\prod_{i=1}^n\prod_{j=1}^m\bigl[1-\hat{\sigma}_{i,j}(u,u')\bigr]
   \Bigr\}\prod dudu'\Bigm|_{J=J'=0}
\end{eqnarray}

\noindent Here $\beta$ is the impact parameter between the two
interacting hadrons $A$ and $B$ and $\hat{\sigma}_{i,j}$ is the
probability for the parton $i$ (of $A$) to have an hard
interaction with the parton $j$ (of $B$).

\noindent The cross section is
obtained by summing all contributions due to all different
hadronic configurations (the sums over $n$ and $m$). For each pair
of values $n$ and $m$, one has a contribution to $\sigma_{hard}$
when at least one hard interaction takes place (whose probability
is represented by the term in curly brackets in the
equation above).

\noindent The cross section, which is analogous to the expression
of the inelastic nucleus-nucleus cross section in the Glauber
model\cite{Bialas:1976ed}, takes into account both disconnected
interactions (which imply $n=m$) and rescatterings (when $n\neq
m$).

\noindent A much simpler expression may be obtained by neglecting
all rescatterings. To that purpose one may write the term in curly
brackets as:

\begin{eqnarray}S\equiv 1-{\rm exp}\sum_{ij}{\rm ln}(1-\hat\sigma_{ij})
=1-{\rm exp}\biggl[-\sum_{ij}\Bigl(\hat\sigma_{ij}+{1\over
2}\hat\sigma_{ij}\hat\sigma_{ij} +\dots\Bigr)\biggr]\end{eqnarray}

\noindent rescatterings correspond to repeated indices. Only the first term of the expansion of the logarithm hence contributes and all rescatterings are removed by the substitution:

\begin{eqnarray}S\Rightarrow 1-{\rm exp}\sum_{ij}\bigl(\hat\sigma_{ij}\bigr)\Rightarrow
  \sum_{ij}\hat\sigma_{ij}-{1\over 2}
  \sum_{ij}\sum_{k\not=i,l\not=j}\hat\sigma_{ij}\hat\sigma_{kl}
  \dots\end{eqnarray}

\noindent The resulting cross section is expressed in a compact way

\begin{eqnarray}\sigma_{hard}(\beta)&=&{\rm exp}(\partial)\cdot {\rm exp}(\partial')
  \Bigl[ 1-{\rm exp}\bigl(-\partial\cdot\hat{\sigma}\cdot\partial'\bigr)\Bigr]
  {\cal Z}_A[J]{\cal Z}_B[J']\Bigm|_{J=J'=0}\cr
  &=&\Bigl[ 1-{\rm exp}\bigl(-\partial\cdot\hat{\sigma}\cdot\partial'\bigr)\Bigr]
  {\cal Z}_A[J]{\cal Z}_B[J']\Bigm|_{J=J'=1}\end{eqnarray}

\noindent where all convolutions are understood. In the simplest non-trivial case all correlations
$C_n$ with $n>2$ can be neglected and the cross section is more explicitly expressed by:

\begin{eqnarray}\sigma_{hard}(\beta)=
  \Bigl[ 1&-&{\rm exp}\Bigl\{{-\int dudu'{\partial_J}\hat{\sigma}(u,u')
  {\partial_{J'}}}\Bigr\}\Bigr]\cr&\cdot&
  {\rm exp}\biggl\{\int D_A(u)J(u)du+{1\over 2}\int C_A(u,v)J(u)J(v)
         dudv\biggr\}\cr
  &\cdot &{\rm exp}\biggl\{\int D_B(u)J(u)du+{1\over 2}\int C_B(u,v)J(u)J(v)
         dudv\biggr\}
\biggm|_{J=J'=0}\end{eqnarray}

\noindent which can be worked out explicitly. One in fact
obtains\cite{Calucci:1991qq}\cite{Calucci:1997ii}

\begin{eqnarray}\sigma_{hard}(\beta)=1-{\rm exp}\Bigl[-{1\over 2}\sum_na_n-{1\over 2}\sum_nb_n/n\Bigr]
\end{eqnarray}

\noindent where

\begin{eqnarray}a_n=(-1)^{n+1}\int&& D_A(u_1)\hat{\sigma}(u_1,u_1')C_B(u_1',u_2')
       \hat{\sigma}(u_2',u_2)
       C_A(u_2,u_3)\dots\cr
       &&\qquad\cdots
       \hat{\sigma}(u_n,u_n')D_B(u_n')
  \prod_{i=1}^ndu_idu_i'\end{eqnarray}

\noindent and

\begin{eqnarray}b_n=(-1)^{n+1}\int &&C_A(u_n,u_1)\hat{\sigma}(u_1,u_1')C_B(u_1',u_2')
           \dots\cr&&\qquad\cdots C_B(u_{n-1}',u_n')\hat{\sigma}(u_n',u_n)
       \prod_{i=1}^ndu_idu_i'\end{eqnarray}

Notice that if one works out the moments of the distribution in
the number of collisions one obtains the same result of the
simplest Poissonian model. From Eq.14 one may express the hard
cross section as a sum of MPI:

\begin{eqnarray}\sigma_{hard}(\beta)&=&\Bigl[ 1-{\rm exp}\bigl(-\partial\cdot\hat{\sigma}\cdot\partial'\bigr)\Bigr]
  {\cal Z}_A[J]{\cal Z}_B[J']\Bigm|_{J=J'=1}\cr
  &=&\sum_{N=1}^{\infty}{\bigl(\partial\cdot\hat{\sigma}\cdot\partial'\bigr)^N\over{N !} } {\rm e}^{-\partial\cdot\hat{\sigma}\cdot\partial'}
  {\cal Z}_A[J]{\cal Z}_B[J']\Bigm|_{J=J'=1}
\end{eqnarray}

\noindent The average number of collisions hence is

\begin{eqnarray}\langle N\rangle\sigma_{hard}(\beta)  &=&\sum_{N=1}^{\infty}{N\bigl(\partial\cdot\hat{\sigma}\cdot\partial'\bigr)^{N}\over{N !} } {\rm e}^{-\partial\cdot\hat{\sigma}\cdot\partial'}
  {\cal Z}_A[J]{\cal Z}_B[J']\Bigm|_{J=J'=1}\cr
  &=&\partial_{J_1}\cdot\hat{\sigma}\cdot\partial_{J_1'}\sum_{N=0}^{\infty}{\bigl(\partial\cdot\hat{\sigma}\cdot\partial'\bigr)^{N}\over{N !} } {\rm e}^{-\partial\cdot\hat{\sigma}\cdot\partial'}
  {\cal Z}_A[J]{\cal Z}_B[J']\Bigm|_{J=J'=1}\cr
  &=&\bigl(\partial_{J_1}\cdot\hat{\sigma}\cdot\partial_{J_1'} \bigr){\cal Z}_A[J]{\cal Z}_B[J']\Bigm|_{J=J'=1}\cr
  &=&\int D_A(x_1;b_1)\hat{\sigma}(x_1x_1') D_B(x_1';b_1-\beta)dx_1dx_1'd^2b_1\equiv\sigma_S(\beta)
  \end{eqnarray}

  \noindent where now $\hat{\sigma}(x_1x_1')$ is the parton-parton cross section integrated with $p_t>p_t^c$ since, given the localization of the interactions in transverse space,  the parton-parton interaction probability has been treated as a $\delta$ as a function of the transverse coordinates: $\hat{\sigma}(u,u')=\hat{\sigma}(x, x')\delta({\bf b}-{\bf b}')$. Analogously one obtains

\begin{eqnarray}{\langle N(N-1)\rangle\over2!}\sigma_{hard}(\beta)&=&{1\over2!}\int D_A(x_1x_2;b_1 b_2)\hat{\sigma}(x_1x_1')\hat{\sigma}(x_2x_2')\nonumber\\&&\quad\quad\times   D_B(x_1' x_2';b_1-\beta, b_2-\beta)dx_1dx_1'd^2b_1 dx_2dx_2'd^2b_2\nonumber\\
&\equiv&\sigma_D(\beta)
\end{eqnarray}

\noindent and in general

\begin{eqnarray}&&\!\!\!\!\!\!\!\!\!\!\!{\langle N(N-1)\dots(N-K+1)\rangle\over K!}\sigma_{hard}(\beta)\cr
  &&\qquad\qquad={1\over K!}\int D_A(x_1 \dots x_K;b_1 \dots b_K)\hat{\sigma}(x_1x_1')\dots\hat{\sigma}(x_Kx_K')\nonumber\\&&\qquad\quad\qquad\quad\times  D_B(x_1' \dots x_K';b_1-\beta \dots b_K-\beta)dx_1dx_1'd^2b_1\dots dx_Kdx_K'd^2b_K \nonumber\\
&&\qquad\qquad\equiv\sigma_K(\beta)
\end{eqnarray}

\noindent which hence shows that {\it when rescatterings are
neglected}, for any choice of multiparton distributions, the
inclusive cross sections are given by the moments of the
distribution in the number of collisions.

\section{"Exclusive" cross sections}

As discussed in the previous section, the inclusive cross
sections are basically the moments of the distribution of the
number of collisions. The moments contain the whole information on
the distribution in the number of collisions ordered in
complexity: The average number is the most basic information on
the distribution and is given by the single scattering inclusive
cross section of the QCD parton model which, precisely for this reason,
represents also the simplest theoretical quantity to evaluate. Its
expression is in fact particularly simple, the convolution of the
parton distributions (which represent the average number of partons in
the hadron in a given kinematical configuration) and the single scattering
partonic cross section.

Analogously the $K$-parton scattering inclusive cross section
gives the $K$th moment of the distribution in the number of
collisions and is related directly to the $K$-body parton distribution of the hadron structure.

As already pointed out, a way alternative to the set of moments, to
provide the information of the distribution, is represented by the set of the different
terms of the probability distribution of multiple
collisions. Correspondingly, in addition to the set of the inclusive cross sections, one may consider the set of the "exclusive" cross sections.
As discussed in the previous section the case where only
two-body parton correlations are taken into account can be worked
out in details. In the following we will hence work out explicit
expression for the single and for the double parton scattering
"exclusive" cross sections.

To obtain the distribution of multiple interactions one may start
from the partonic interaction probability

\begin{eqnarray}1-\prod_{i,j=1}^n(1-\hat\sigma_{ij})
\end{eqnarray}

\noindent where in the product each index assumes a given value
only once, in such a way that possible re-interactions are not
included. The probability of a single interaction may hence be
expressed as

\begin{eqnarray}\Biggl(-{\partial\over\partial g}\Biggr)\prod_{i,j=1}^n(1-g\hat\sigma_{ij})\Bigg|_{g=1}=\sum_{kl}\hat\sigma_{kl}\prod_{ij\neq kl}^n(1-g\hat\sigma_{ij})\Bigg|_{g=1}\end{eqnarray}

\noindent while the probability for a double collisions is

\begin{eqnarray}{1\over 2!}\Biggl(-{\partial\over\partial g}\Biggr)^2\prod_{i,j=1}^n(1-g\hat\sigma_{ij})\Bigg|_{g=1}={1\over 2!}\sum_{kl}\sum_{rs}\hat\sigma_{kl}\hat\sigma_{rs}\prod_{ij\neq kl,rs}^n(1-g\hat\sigma_{ij})\Bigg|_{g=1}\end{eqnarray}

\noindent The expressions for the single and double "exclusive"
integrated cross sections are

\begin{eqnarray}\Biggl(-{\partial\over\partial
g}\Biggr)\esp{-X(g)}\Bigg|_{g=1}&=&X'(g)\esp{-X(g)}\Bigg|_{g=1}\\
{1\over2!}\Biggl(-{\partial\over\partial
g}\Biggr)^2\esp{-X(g)}\Bigg|_{g=1}&=&{1\over2!}\Bigl\{[X'(g)]^2-X''(g)\Bigr\}\esp{-X(g)}\Bigg|_{g=1}
\end{eqnarray}

\noindent where $X={1\over2}(\sum a_n+\sum b_n/n)$ with $a_n$ and $b_n$
given by Eq.'s 17 and 18.

The physical description of the process is purely probabilistic. As a consequence, the
hard cross section may be unitarized in each given phase space volume,
which hence defines also all integration limits.

By restricting the rapidity window or the $p_t$ interval one may
control the importance of the terms with different powers of
$\hat\sigma$. In a small window only the terms linear in
$\hat\sigma$ are relevant. It is hence meaningful to expand the
terms above in powers of $\hat\sigma$. Going up to order
$\hat\sigma^2$ one obtains for $X'$ and $X''$

\begin{eqnarray}
X'(u,u')&=&D_A(u)\hat{\sigma}(u,u') D_B(u')\nonumber\\
&&\qquad-\Bigl[\int D_A(u)\hat{\sigma}(u,u') C_B(u',u_1')
       \hat{\sigma}(u_1',u_1)D_A(u_1)du_1du_1'
+A\leftrightarrow B\Bigr]\nonumber\\
&&\qquad-\int C_A(u_1,u)\hat{\sigma}(u,u') C_B(u',u_1')
       \hat{\sigma}(u_1',u_1)du_1du_1'\\
X''(u_1,u_1',u_2,u_2')&=&-\Bigl[ D_A(u_1)\hat{\sigma}(u_1,u_1')
C_B(u_1',u_2')
       \hat{\sigma}(u_2,u_2')D_A(u_2)
+A\leftrightarrow B\Bigr]\nonumber\\
&&\qquad- C_A(u_1,u_2)\hat{\sigma}(u_1,u_1') C_B(u_1',u_2')
       \hat{\sigma}(u_2,u_2')\end{eqnarray}

\noindent At order $\hat\sigma^2$, the single scattering "exclusive" cross section $\tilde\sigma_1$ is given by

\begin{eqnarray}
\tilde\sigma_1(u,u')&=& D_A(u)\hat{\sigma}(u,u') D_B(u')\Bigl[1- \int D_A(u_1)\hat{\sigma}(u_1,u_1') D_B(u_1')du_1du_1'\Bigr]\nonumber\\
&&\qquad-\Bigl[\int D_A(u)\hat{\sigma}(u,u') C_B(u',u_1')
       \hat{\sigma}(u_1',u_1)D_A(u_1)du_1du_1'
+A\leftrightarrow B\Bigr]\nonumber\\
&&\qquad-\int C_A(u_1,u)\hat{\sigma}(u,u') C_B(u',u_1')
       \hat{\sigma}(u_1',u_1)du_1du_1'
       \end{eqnarray}

\noindent while the expression of the double scattering "exclusive" cross section $\tilde\sigma_2$ is

\begin{eqnarray}
\tilde\sigma_2(u_1,u_1',u_2,u_2')={1\over2!}&\Bigl\{&D_A(u_1)\hat{\sigma}(u_1,u_1') D_B(u_1')\cdot D_A(u_2)\hat{\sigma}(u_2,u_2') D_B(u_2')\nonumber\\
&&+\Bigl[ D_A(u)\hat{\sigma}(u_1,u_1') C_B(u_1',u_2')
       \hat{\sigma}(u_2,u_2')D_A(u_2)
+A\leftrightarrow B\Bigr]\nonumber\\
&&+ C_A(u_1,u_2)\hat{\sigma}(u_1,u_1') C_B(u_1',u_2')
       \hat{\sigma}(u_2,u_2')\Bigr\}\end{eqnarray}

\noindent One may immediately verify that

\begin{eqnarray}
\int\tilde\sigma_1(u,u')dudu'+2!\int\tilde\sigma_2(u_1,u_1',u_2,u_2')du_1du_1'du_2du_2'&=&\int D_A(u)\hat{\sigma}(u,u') D_B(u')dudu'\nonumber\\
&=&\langle N\rangle\sigma_{hard}\equiv\sigma_S
\end{eqnarray}

By restricting the phase space interval of the observed final
state, $\tilde\sigma_1$ is well expressed by the term linear in
$\hat\sigma$, which means that in that limit the probability of
interaction is well approximated by the average of the
distribution, while $\tilde\sigma_2$ is negligibly small. In other
words the single parton scattering "exclusive" cross section is
well represented by the single scattering expression of the simple
QCD parton model. When the phase space volume is increased, the
single parton scattering "exclusive" cross section becomes
increasingly different from the prediction of the QCD parton
model and, as it will be shown in the explicit examples of the next two paragraphs, the difference allows a direct measure of the importance
of correlations. Notice that, by summing the expressions of
$\tilde\sigma_1$ and $\tilde\sigma_2$ at order $\hat\sigma^2$ with
the proper multiplicity factors one obtains, as shown above, the
{\it inclusive} single parton scattering cross section, which is
correctly given by the QCD parton model expression. By comparing
the measured inclusive cross section with a sum of the measured
"exclusive" cross sections, taken with the proper multiplicity
factors, up to a given order in the number of collisions, one
hence has a direct indication of the importance of higher order
unitarity corrections in a given phase space interval.

To show how these ideas may be explicitly implemented we work out two explicit examples in the following paragraphs 4.1 and 4.2.
The transverse parton coordinates are not accessible
experimentally. On the other hand while, at small $x$, partons may
not be correlated in momentum fraction, as conservation
constraints may be diluted when the parton population is large,
they are surely correlated in their transverse coordinates, since
they are all confined inside the same hadron. A simplest
possibility is hence to assume that the only relevant correlations are
those in the transverse coordinates. To have some indication we will consider two different analytically solvable models of the parton densities in transverse space, the Gaussian and the exponential model, while neglecting in our two examples all effects of longitudinal correlations.

\subsection{ Gaussian model}

One may assume Gaussian distributions for the parton
densities and for the correlations:

\begin{eqnarray}
D(x,b)&=&G(x)f(b)\nonumber\\
f(b)&=&g(b,R^2)\nonumber\\
C(x_1,x_2;b_1,b_2)&=&G(x_1)G(x_2)h(b_1,b_2)\nonumber\\
h(b_1,b_2)&=&c\cdot g(B,R^2/2)\bar h(b,\lambda)\nonumber\\
\bar h(b,\lambda)&=&\eta {d\over d\gamma}g(b,\lambda^2/\gamma)\Big|_{\gamma=1} \nonumber\\
g(b,R^2)&=&{1\over\pi R^2}{\rm exp}(-b^2/R^2)
\end{eqnarray}

\noindent where

\begin{eqnarray}
{\bf B}&=&[{\bf b}_1+{\bf b}_2]/2\nonumber\\
{\bf b}&=&[{\bf b}_1-{\bf b}_2]
\end{eqnarray}

\noindent in such a way that the following relations hold

\begin{eqnarray}
&&\int d^2b g(b,R^2)=1,\qquad\int d^2b_2 g({\bf b}_1-{\bf b}_2,R_1^2)g({\bf b}_2,R_2^2)=g({\bf b}_1,R_1^2+R_2^2)
\nonumber\\
&&\int d^2b h(b_1,b_2)d^2b=0
\end{eqnarray}

\noindent The connection with the root mean square hadron radius
$\langle r^2\rangle\equiv\bar R^2$ is $R^2=2\bar R^2/3$. We define
the correlation $r_c$ as the value of $b$ where the function $\bar
h(b,\lambda)$ changes sign. In this case one hence has
$r_c=\lambda$. In order to define unambiguously the correlation
strength we choose the normalization
$$\int_{|b|\le r_c|} \bar h(b,\lambda^2)\, d^2b=1\;,$$
 since the integral over the whole $b$-space gives zero. With this
 choice $\eta=e$, where $e$ is the Euler number.
 The integrations on the transverse variables of the
various terms in Eq.s 30 and 31 give

\begin{eqnarray}
D_A\hat{\sigma}D_B&\rightarrow&\int  d^2b d^2\beta g_A({\bf b}-{\bbeta},R_1^2)g_B({\bf b},R_2^2)=1\nonumber\\
D_A\hat{\sigma}D_B\cdot D_A\hat{\sigma}D_B&\rightarrow&\int  d^2b_1d^2b_2 d^2\beta g_A({\bf b}_1-{\bbeta},R_1^2)g_B({\bf b}_1,R_2^2) g_A({\bf b}_2-{\bf \beta},R_1^2)g_B({\bf b}_2,R_2^2)\nonumber\\&&={1\over2\pi (R_1^2+R_2^2)}\nonumber\\
D_A\hat{\sigma}C_B\hat{\sigma}D_A&\rightarrow&\int  d^2b_1d^2b_2 d^2\beta g_A({\bf b}_1-{\bbeta},R_1^2)h_B({\bf b}_1,{\bf b}_2)g_A({\bf b}_2-{\bbeta},R_1^2)\nonumber\\
&&={c\,e\over\pi}{\lambda^2\over (2R_1^2+\lambda^2)^2}\nonumber\\
C_A\hat{\sigma}C_B\hat{\sigma}&\rightarrow&\int  d^2b_1d^2b_2 d^2\beta h_A({\bf b}_1-{\bbeta},{\bf b}_2-{\bbeta})h_B({\bf b}_1,{\bf b}_2)\nonumber\\
&&={c^2\over4\pi}{e^2\over\lambda^2}
\end{eqnarray}

\noindent the "exclusive" cross sections $\tilde\sigma_1$ and of
$\tilde\sigma_2$ are hence explicitly expressed in this case in
terms of the QCD-parton model single scattering cross section
$\sigma_S$, the transverse hadron size $R$ and the correlations
parameters $c$ and $\lambda$:

\begin{eqnarray}
{d\tilde\sigma_1\over dy d{\bf p}_t}&=&{d\sigma_S\over dy d{\bf p}_t}\times\Bigl[1-{\sigma_S\over4\pi R^2}-c\,e{2\sigma_S\lambda^2\over\pi(2R^2+\lambda^2)^2}-c^2\,e^2{\sigma_S\over4\pi\lambda^2}\Bigr]\nonumber\\&=&{d\sigma_S\over dy d{\bf p}_t}\times\Bigl[1-{\sigma_S\over\sigma_{eff}}\Bigr]\nonumber\\
{d\tilde\sigma_2\over dy_1dy_2 d{\bf p}_{t1}d{\bf
p}_{t2}}&=&{1\over 2}{d\sigma_S\over dy_1 d{\bf
p}_{t1}}{d\sigma_S\over dy_2 d{\bf p}_{t2}}\times\Bigl[{1\over4\pi
R^2}+c\,e{2\lambda^2\over\pi(2R^2+\lambda^2)^2}+c^2 e^2{1\over4\pi\lambda^2}\Bigr]\nonumber\\&=&{1\over
2\sigma_{eff}}{d\sigma_S\over dy_1 d{\bf p}_{t1}}{d\sigma_S\over
dy_2 d{\bf p}_{t2}}\equiv{d\sigma_D\over dy_1dy_2 d{\bf
p}_{t1}d{\bf p}_{t2}}\end{eqnarray}

\noindent where $\sigma_{eff}$ is the scale factor which
characterizes the inclusive rate of double parton collisions. As a
function of the root mean hadron radius $\bar R$, of the
correlation length $r_c$ and of the correlation strength $c$, the
expression is

\begin{eqnarray}
{1\over\sigma_{eff}}={3\over8\pi \bar R^2}\Bigl\{1+c\,e\cdot16
{3s^2\over(4+3s^2)^2}+c^2\,e^2\cdot{2\over3s^2}\Bigr\}
\end{eqnarray}

\noindent where $s\equiv r_c/\bar R$. Notice that neglecting correlation in fractional
momenta, in each phase space interval $\tilde\sigma_1$ is different from $\sigma_S$ only by the
overall normalization (related to $\sigma_{eff}$ as shown above). At this order in $\hat\sigma$,
$\tilde\sigma_2$ coincides with the expression of the inclusive
double parton scattering cross section while, in this case, the
scale factor $\sigma_{eff}$ does not depend on $x$.

\subsection{Exponential model}

A different option is to assume an exponential for the three dimensional parton density

\begin{eqnarray}
\rho(r)&=&{\mu^3\over8\pi}\esp{-\mu r},\qquad \int\rho(r)d^3r=1\nonumber\\
D(x,b)&=&G(x)f(b)\nonumber\\
f(b)&=&\int\rho(r)dz={1\over4\pi}\mu^3bK_1(\mu b)\nonumber\\
\end{eqnarray}

\noindent where ${\bf b}$ is the transverse component of ${\bf r}$ and $K_1$ is the modified Bessel function of the second kind.
It may be seen that the two distributions Gaussian and exponential are
   formally connected. If one performs a linear combination of the Gaussian
   distributions (33) with appropriate weight one ends up with the expression (39):

\begin{eqnarray}
\int_0^{\infty}{\mu^4\over2(2\lambda)^3}\esp{-\mu^2/4\lambda}g(b,1/\lambda)d\lambda={1\over4\pi}\mu^3bK_1(\mu b)
\end{eqnarray}

For the next calculations it is convenient to take the Fourier transform of the transverse parton density

\begin{eqnarray}
\rho(b)&=&{1\over(2\pi)^2}\int\esp{-i{\bf p_t}\cdot{\bf b}}\tilde\rho(p_t)d^2 p_t\nonumber\\
\tilde\rho(p_t)&=&{\mu^4\over(\mu^2+p_t^2)^2}\nonumber\\
\end{eqnarray}

\noindent One may assume for the correlation the expression

\begin{eqnarray}
C(x_1,x_2;b_1,b_2)&=&G(x_1)G(x_2)h(b_1,b_2)\nonumber\\
h(b_1,b_2)&=&c\cdot f(B)\bar h(b)\nonumber\\
\bar h(b)&=&{1\over(2\pi)^2}\int\esp{-i{\bf p}_t\cdot{\bf b}}\tilde h(p_t)d^2p_t\nonumber\\
\tilde h(p_t)&=&\eta\frac {d}{d\gamma}\frac
{\kappa^4\gamma^2}{(\kappa^2+p_t^2)^2}\Big|_{\gamma=1}
=2\eta{{\kappa^4 p_t^2}\over{(p_t^2+\kappa^2)^3}}
\end{eqnarray}

\noindent in such a way that

\begin{eqnarray}
\tilde h(0)=\int\bar h(b)d^2b=0
\end{eqnarray}

\noindent As a function of the coordinates the expression is

\begin{eqnarray}
\bar h(b)={\eta\over2\pi}\Bigl\{\kappa^3b K_1(\kappa
b)-{1\over4}\kappa^4 b^2K_2(\kappa b)\Bigr\}
\end{eqnarray}

\noindent where $K_1$ and $K_2$ are modified Bessel functions. The
root mean square value correlation term is performed in the same way as in  of the density is $\langle r^2\rangle\equiv
\bar R^2=12/\mu^2$ and, like in the previous case, defining the
correlation length as the value of $b$ where the function $\bar
h(b)$ changes sign, one has $r_c=x_0/\kappa$, where
$x_0\simeq2.386$ solves the equation

\begin{eqnarray}
4x_0K_1(x_0)=x_0^2K_2(x_0)
\end{eqnarray}
\noindent The normalization of the correlation function is performed in the same way as in the Gaussian model, it gives now $\eta=3.456$.
 The integrations on the transverse variables of the
various terms in Eq.s 30 and 31 give

\begin{eqnarray}
D_A\hat{\sigma}D_B&\rightarrow&\int  d^2b d^2\beta f_A({\bf b}-{\bbeta})f_B({\bf b})=1\nonumber\\
D_A\hat{\sigma}D_B\cdot D_A\hat{\sigma}D_B&\rightarrow&\int  d^2b_1d^2b_2 d^2\beta f_A({\bf b}_1-{\bbeta})f_B({\bf b}_1) f_A({\bf b}_2-{\bf \beta})f_B({\bf b}_2)\nonumber\\
&&={3\over7\pi \bar R^2}\nonumber\\
D_A\hat{\sigma}C_B\hat{\sigma}D_A&\rightarrow&\int  d^2b_1d^2b_2 d^2\beta f_A({\bf b}_1-{\bbeta})\bar h_B({\bf b}_1,{\bf b}_2)f_A({\bf b}_2-{\bbeta})\nonumber\\
&&=c{3\eta\over\pi \bar R^2}F\Bigl[{x_0^2\bar R^2\over12r_c^2}\Bigr]\nonumber\\
F(a)&\equiv&{3+44a-36a^2-12a^3+a^4+12a(2+3a){\rm ln}(a)\over 3(a-1)^6a}\nonumber\\
C_A\hat{\sigma}C_B\hat{\sigma}&\rightarrow&\int  d^2b_1d^2b_2 d^2\beta \bar h_A({\bf b}_1-{\bbeta},{\bf b}_2-{\bbeta})\bar h_B({\bf b}_1,{\bf b}_2)\nonumber\\
&&=c^2\,\eta^2{1\over 30\pi}{x_0^2\over r_c^2}
\end{eqnarray}

\noindent Notice that the function $F(a)$ is regular near $a=1$,
the expansion being

\begin{eqnarray}
F(a)={1\over 15}-{a-1\over 7}+3{(a-1)^2\over 14}-5{(a-1)^3\over 18}+{\cal
O}\Bigl((a-1)^4\Bigr)
\end{eqnarray}

\noindent As a function of the root mean square hadron radius
$\bar R$, of the ratio $s\equiv r_c/\bar R$ and of the correlation
strength $c$, the expression of the effective cross section is

\begin{eqnarray}
{1\over\sigma_{eff}}={3\over7\pi \bar R^2}\Bigl\{1+c\,\eta\cdot14\cdot
F\Bigl({1\over12}{x_0^2\over
s^2}\Bigr)+c^2\,\eta^2\cdot{7\over30}\Bigl({1\over 3}{x_0^2\over
s^2}\Bigr)\Bigr\}
\end{eqnarray}

One may hence obtain information on the correlations by
comparing the measured "exclusive" single scattering cross section
with the measured inclusive single scattering cross section:

\begin{eqnarray}
{d\tilde\sigma_1\over dy d{\bf p}_t}-{d\sigma_S\over dy d{\bf
p}_t}={d\sigma_S\over dy d{\bf p}_t}{\sigma_S\over\sigma_{eff}}
\end{eqnarray}

\noindent The cases considered here are relevant for a phase space window where multiple
collisions, of order higher than two, do not give important contributions; condition which may be always achieved by taking, for example, the lower cutoff in $p_t$ large enough. Notice that the importance of higher order scattering terms, namely of unitarity corrections, is controlled irrespectively of any simplifying hypothesis on correlations, by verifying the validity of the sum rule Eq.32.

\noindent The relation above
allows to obtain the value of $\sigma_{eff}$. By subtracting
from the effective cross section the term ${3/(7\pi \bar R^2)}$, or
${3/(8\pi \bar R^2)}$, depending on the prejudice on the parton
density in transverse space, one obtains a direct information on
correlations. Notice that, although the two options considered,
the Gaussian and the exponential, correspond to rather different parton densities, the
quantity to be subtracted is basically the same in the two cases
and it is essentially determined only by the value of the
mean square hadron radius $\bar R^2$. Indications on the value of $\bar
R^2$ are obtained by the measurement of the generalized parton
distributions\cite{Mueller:1998fv}\cite{Belitsky:2005qn}. In the
case of gluons one may estimate $\bar R^2\simeq.42$
fm$^2$\cite{Frankfurt:2003td}.

One should point out that the validity of Eq.49 is easily tested,
since the dependence of the difference in the right hand side  of
the equation, integrated in a given phase space interval, on the
rapidity window and on the lower cutoff in $p_t$ has to be the
same as the dependence of $\sigma_S^2$, integrated in the same
phase space interval. So, if the sum rule of Eq.32 is satisfied
while the relation in Eq.49 is not, one may safely conclude that
one has direct evidence of longitudinal correlations in the hadron
structure.

The analysis may be repeated by including higher order multiple
scattering terms in Eq.30 and in the following. In such a case,
one may write relations similar to Eq.49, linking higher order
multiple scattering inclusive cross sections with various
combinations of "exclusive" cross sections. Independent
constraints on the correlation parameters may hence be
established, hopefully allowing an unambiguous identification of
the correlation parameters.

\section{Concluding remarks}

MPI are originated by the large flux of
partons in high energy hadronic collisions. The dominant
contribution to the process is hence due to the terms which
maximize the incoming parton flux. Given the number of partonic
collisions, the incoming flux is maximized by the set of terms
where elementary hard interactions are all disconnected one from
another or, more precisely, where the connection is only due to
soft exchanges. Being the interactions uncorrelated with respect
to hard dynamics, all correlations observable in the final state
are originated by the initial state flux, namely by the hadron
structure. In such a scheme all partonic collisions add
incoherently in the final cross section, leading to a purely
probabilistic picture of the process. It is hence natural to
identify two different classes of hard cross sections, the
inclusive cross section, related to the moments of the
distribution, and the "exclusive" cross sections, related to the
different terms of the probability distribution in the number of
hard collisions.

A relevant property is that the picture of the interaction is
probabilistic in each phase space interval, in such a way that one
may associate a different probability distributions to each
different phase space choice to observe the final state. Including the
possibility to select events which are fully or only partially
contained in the given phase space interval. Any choice identifies
an interval in $p_t$ and restricts in a precise way the limits in
$x$ of the initial state, which identifies the domain of
definition of the probability distribution of multiple collisions.
One should point out that the same restricted intervals define
also the integration range of the integrated terms, which appear
in the "exclusive" differential cross sections (e.g. the
integrations in Eq.30). Integrated terms appear in the "exclusive"
differential cross sections because of the normalization of the
probability distribution, their origin may be traced back to the
virtual corrections of the process. Normalization hence fixes
unambiguously the integration limits of the virtual terms, which
must coincide with the kinematical limits imposed to the real
terms by the choice adopted to select the final state.

Another relevant point is that, as proved in the appendix, the
components of the hadron structure, namely the terms $D$ and
$C_n$, are well defined quantities in the picture of the
interaction, as they do not mix when the selection choice of the
final state is modified. The effect, to modify the selection
choice of the final state, is in fact only to change the
integration domain of each term, namely to probe the multi-parton
structure of the hadron in different domains in $x$ and $Q^2$ .

As it was pointed out, the inclusive cross sections can provide a
direct information on the correlations of the multi-parton
structure of the hadron, while to evaluate the "exclusive" cross
sections one needs, in principle, an infinite non perturbative
information. The expansion of the "exclusive" cross sections in
the number of parton collisions allows however to identify the
dominant terms and to isolate the leading contributions due to the
presence of correlations. Explicit relations between inclusive and
exclusive cross sections may hence be written at each order of the
expansion in the number of collisions, while the importance of
higher order terms may be controlled by selecting the phase space
interval where to observe the final state. The comparison between
experimentally measured inclusive and "exclusive" cross sections,
at a given order in the number of collisions, provides a model
independent tool to judge on the importance of higher order
unitarity corrections (namely by checking the validity of sum
rules like Eq.32).

In the present paper we have worked out in details, at the lowest
non trivial order, the simplest case where only the correlations
in the transverse coordinates are taken into account. Of course
correlations will depend also on fractional momenta and on the
different kinds of partons considered (gluons and quarks,
distinguishing different flavors and the valence from the sea). To
some extent the different kinds of partons involved may be selected
by looking at definite final states produced and at the
corresponding values of momentum fraction involved (a prompt
photon plus five jets at low $x$ are likely to be generated by a
triple scattering involving a quark and five gluons etc.). The
information on correlations will hence to be related to the
different kinds of initial state partons involved in the
interaction.

While the importance of higher order unitarity corrections is
tested, in a given phase space interval, by verifying the sum rule
in Eq.32 (and similar when including a higher number of
collisions) terms with correlations are isolated by using
relations like Eq.49, which allow to obtain the value of the
effective cross section and of similar scale factors for higher
orders unitarity corrections. The importance of correlations is
tested by comparing with the expectation in the uncorrelated case
and, in particular, longitudinal momentum correlations will show
up with the dependence of the effective cross section (and of the
similar scale factors for higher orders unitarity corrections) on
the rapidity window. With respect to longitudinal correlations one
should anyhow point out that, while indications on the effects of
longitudinal correlations may be obtained along these lines in
$pp$ collisions, a model independent separation of longitudinal
and transverse correlations may be obtained by looking at MPI in
the case of hadron-nucleus collisions\cite{Strikman:2001gz}.

\begin{acknowledgments}
One of us (D.T.) thanks Mark Strikman and Ted Rogers for useful
discussions.
\end{acknowledgments}

\begin{center}
    {\bf APPENDIX}
 \end{center}

As already mentioned in the main text the "exclusive" cross
sections are defined with respect to intervals of variation of the
kinematical variables, which are to some extent arbitrary. Even
though the result may look intuitive, we like to analyze formally
what happens when varying the intervals of the kinematical
variables: Starting with the formalism stated in eq. (8) we call
$\cal U$ the field of variation of $u$ and divide it into two
mutually exclusive parts $\cal U=\cal R\cup\cal S$.
Correspondingly it is convenient to denote by $r$ the variable $u$
when if varies within $\cal R$ and $s$ when it varies within $S$.
Then for the exclusive distribution within $\cal R$ we have:
$$\tilde W_1(r_1)=W_1(r_1)+\int_{\cal S} W_2(r_1,s)ds+\frac {1} {2}\int_{\cal S} W_3(r_1,s,s')ds\;ds'+\frac {1}  {3!}\int_{\cal S} W_4(r_1,s,s',s")ds\;ds'\;ds''+\dots$$

$$\tilde W_2(r_1,r_2)= W_2(r_1,r_2)+\int_{\cal S} W_3(r_1,r_2,s)ds+\frac {1}  {2}\int_{\cal S} W_4(r_1,r_2,s,s')ds\;ds'+\dots$$
These relations may be summarized by means of the following
procedure: Let us define the projectors $R(u)=1, S(u)=0$ for
$u\in\cal R$ $R(u)=0, S(u)=1$ for $u\in\cal S$, $R+S=1$; and the
generating functional $\tilde {\cal Z}[J]= {\cal Z}[R\cdot J+S]$.
Then expanding $\tilde {\cal Z}[J]$ as in eq (8) and using the
symmetry of the $W_n$ it is easy to attain the result:
$$\tilde {\cal Z}[J]=\sum_n\int_{\cal R}J(r_1)\dots J(r_n)\tilde W_n(r_1\dots r_n) dr_1\dots dr_n$$
Now we perform the same substitution in the logarithm of the
generating functional $\tilde {\cal F}[J]={\cal F}[R\cdot J+S]$
and using $1-S=R$ the projector $R$ is found in front of every
integrand, so the overall effect is a restriction in the
integration domain and the equivalent of eq. (10) is easily found:
$$
\tilde {\cal F}[J]=\int_{\cal R}
D_1(r)[J(r)-1]du+\sum_{n=2}^{\infty} \frac {1}{ n!} \int_{\cal R}
C_n(r_1\dots r_n)\bigl[J(r_1)-1\bigr]\dots \bigl[J(r_n)-1\bigr]
dr_1\dots dr_n $$

The conclusion is hence that the inclusive quantities are not modified by the
variation of the interval where the kinematical variables are defined, on the contrary the variation of the
exclusive quantities can be worked out. With respect to the previous
formula the difference arises only from the different
integration domain of the terms $D$ and $C_n$.

\end{document}